\documentclass[letterpaper,english,aps,letter,prl,twocolumn,superscriptaddress]{revtex4}
\usepackage[T1]{fontenc}
\usepackage[latin1]{inputenc}
\usepackage{xcolor}
\usepackage{pdfcolmk}
\usepackage{babel}

\usepackage{amsmath}
\usepackage{graphicx}
\usepackage{amssymb}
\PassOptionsToPackage{normalem}{ulem}
\usepackage{ulem}
\usepackage[unicode=true, pdfusetitle,
 bookmarks=true,bookmarksnumbered=false,bookmarksopen=false,
 breaklinks=false,pdfborder={0 0 0},backref=false,colorlinks=false]
 {hyperref}

\makeatletter

\pdfpageheight\paperheight
\pdfpagewidth\paperwidth

\providecolor{lyxadded}{rgb}{1,0,0}
\providecolor{lyxdeleted}{rgb}{0,0,1}

\@ifundefined{textcolor}{}
{%
 \definecolor{BLACK}{gray}{0}
 \definecolor{WHITE}{gray}{1}
 \definecolor{RED}{rgb}{1,0,0}
 \definecolor{GREEN}{rgb}{0,1,0}
 \definecolor{BLUE}{rgb}{0,0,1}
 \definecolor{CYAN}{cmyk}{1,0,0,0}
 \definecolor{MAGENTA}{cmyk}{0,1,0,0}
 \definecolor{YELLOW}{cmyk}{0,0,1,0}
 }

\usepackage{ae,aecompl}
\renewcommand{\citet}[1]{\cite{#1}}

\makeatother

\begin{document}
\setlength{\abovedisplayskip}{0.4ex}\setlength{\belowdisplayskip}{0.4ex}
\setlength{\abovedisplayshortskip}{0.15ex}\setlength{\belowdisplayshortskip}{0.15ex}
\setlength{\belowcaptionskip}{0mm} \setlength{\abovecaptionskip}{0mm}
\setlength{\textfloatsep}{1mm}

\global\long\def\V#1{\boldsymbol{#1}}
\global\long\def\M#1{\boldsymbol{#1}}
\global\long\def\Set#1{\mathbb{#1}}

\global\long\def\D#1{\Delta#1}
\global\long\def\d#1{\delta#1}

\global\long\def\norm#1{\left\Vert #1\right\Vert }
\global\long\def\abs#1{\left|#1\right|}

\global\long\def\grad{\M{\nabla}}
\global\long\def\Div{\M{\nabla}\cdot}

\global\long\def\avv#1{\left\langle #1\right\rangle }
\global\long\def\av#1{\langle#1\rangle}

\global\long\def\ki{k}
\global\long\def\wi{\omega}

\title{\emph{Diffusive Transport Enhanced by Thermal Velocity Fluctuations}}

\author{Aleksandar Donev}

\email{donev@courant.nyu.edu}

\affiliation{Courant Institute of Mathematical Sciences, New York University,
New York, NY 10012}

\author{John B. Bell}

\affiliation{Center for Computational Science and Engineering, Lawrence Berkeley
National Laboratory, Berkeley, CA, 94720}

\author{Anton de la Fuente}

\affiliation{Department of Physics, San Jose State University, San Jose, California,
95192}

\author{Alejandro L. Garcia}

\affiliation{Department of Physics, San Jose State University, San Jose, California,
95192}
\begin{abstract}
We study the contribution of advection by thermal velocity fluctuations
to the effective diffusion coefficient in a mixture of two indistinguishable
fluids. We find good agreement between a simple fluctuating hydrodynamics
theory and particle and finite-volume simulations. The enhancement
of the diffusive transport depends on the system size $L$ and grows
as $\ln(L/L_{0})$ in quasi two-dimensional systems, while in three
dimensions it scales as $L_{0}^{-1}-L^{-1}$, where $L_{0}$ is a
reference length. Our results demonstrate that fluctuations play an
important role in the hydrodynamics of small-scale systems.
\end{abstract}
\maketitle
Thermal fluctuations in non-equilibrium systems in which a constant
(temperature, concentration, velocity) gradient is imposed externally
exhibit remarkable behavior compared to equilibrium systems \citet{FluctHydroNonEq_Book}.
The solution of the linearized equations of fluctuating hydrodynamics
shows that concentration and density fluctuations exhibit long-ranged
correlations in the presence of a macroscopic concentration gradient
$\grad c$ \citet{LongRangeCorrelations_MD,ExtraDiffusion_Vailati,FluctHydroNonEq_Book}.
The enhancement of large-scale (small wavenumber) concentration fluctuations
is dramatic during the early stages of diffusive mixing between initially
phase-separated fluids. These \emph{giant fluctuations }\citet{GiantFluctuations_Nature,GiantFluctuations_Cannell,GiantFluctuations_ThinFilms}
during free diffusive mixing have been measured using light scattering
and shadowgraphy techniques \citet{GiantFluctuations_Nature,GiantFluctuations_Cannell,FractalDiffusion_Microgravity},
finding good but imperfect agreement with theoretical predictions.

The giant fluctuation phenomenon arises because of the appearance
of long-ranged correlations between concentration and velocity fluctuations
in the presence of a concentration gradient. It has been predicted
that these correlations give rise to fluctuation-renormalized transport
coefficients \citet{DiffusionRenormalization_I,ExtraDiffusion_Vailati};
however, the predicted enhancement of transport at hydrodynamic scales
has not yet been computationally observed. In particular, it is important
to understand how the effective transport coefficients depend on the
length scale of observation.

In this Letter we consider diffusion in\emph{ }a mixture of identical
but labeled (as components 1 and 2) fluids \citet{SelfDiffusion_Linearity}
enclosed in a box of size $L_{x}\times L_{y}\times L_{z}$, in the
absence of gravity. Periodic boundary conditions are applied in the
$x$ (horizontal) and $z$ (depth) directions, while the top and bottom
boundaries are impermeable constant-temperature walls. A concentration
gradient $\nabla\bar{c}=(c_{T}-c_{B})/L_{y}$ is imposed along the
$y$ axes by enforcing a constant concentration $c_{T}$ at the top
wall and $c_{B}$ at the bottom wall. Because the fluids are indistinguishable,
concentration is \emph{passively }transported by thermal fluctuations.

Since species are not changed in particle collisions, the diffusive
transport of concentration $c=\rho_{1}/\rho$ can only occur via advective
motion of the particles, where $\rho$ denotes the mass density. The
mass flux for a given species is therefore equal to the momentum density
for particles of that species. At steady state the particles of a
given species have a non-zero macroscopic momentum density $\bar{\V j}_{1}=\bar{\rho}_{1}\bar{\V v}_{1}=-\bar{\rho}\chi\left(\grad\bar{c}\right)$,
where $\chi$ is the mass diffusion coefficient \citet{BinaryMixKineticTheory}.
The local fluctuations around the macroscopic mean, $\rho_{1}=\bar{\rho}_{1}+\d{\rho}_{1}$
and $\V v_{1}=\bar{\V v}_{1}+\d{\V v}_{1}$, can also make a non-trivial
contribution to the average mass flux if they are correlated,\begin{equation}
\av{\V j_{1}}=\av{\rho_{1}\V v_{1}}=-\bar{\rho}\chi\left(\grad\bar{c}\right)+\av{\left(\d{\rho}_{1}\right)\left(\d{\V v}_{1}\right)}.\label{eq:av_j1}\end{equation}

At mesoscopic scales the hydrodynamic behavior of fluids can be described
with the Landau-Lifshitz Navier-Stokes (LLNS) equations of fluctuating
hydrodynamics \citet{Landau:Fluid,FluctHydroNonEq_Book}. The incompressible
isothermal LLNS equations for a mixture of two indistinguishable fluids
are \begin{align}
\partial_{t}\V v & \!=-\grad\pi-\V v\!\cdot\!\grad\V v+\nu\grad^{2}\V v+\grad\!\cdot\!(A_{v}\,\M{\mathcal{W}}),\label{eq:LLNS_incomp_v}\\
\partial_{t}c\! & =\!-\V v\!\cdot\!\grad c+\chi\grad^{2}c+\grad\!\cdot\!(A_{c}\,\widetilde{\M{\mathcal{W}}})\label{eq:LLNS_incomp_c}\end{align}
where $\eta$ is the viscosity and $\nu=\eta/\rho$, and the pressure
$\pi$ enforces $\grad\!\cdot\!\V v=0$. The stochastic fluxes are
white-noise random Gaussian tensor $\M{\mathcal{W}}$ and vector $\widetilde{\M{\mathcal{W}}}$
fields, with amplitudes $A_{v}^{2}=2\eta k_{B}T/\rho^{2}$ and $A_{c}^{2}=2m\chi c(1-c)/\rho$
determined from the fluctuation-dissipation principle, where $m$
is the fluid particle mass.

In addition to the usual Fickian contribution, the diffusive flux
in (\ref{eq:LLNS_incomp_c}) includes advection by the fluctuating
velocities $\V v=\d{\V v}$,\[
-\V v\cdot\grad\left(\d c\right)+\chi\grad^{2}\left(\delta c\right)=\grad\cdot\left[-\left(\delta c\right)\left(\d{\V v}\right)+\chi\grad\left(\delta c\right)\right],\]
which is a quadratic function of the fluctuations. To leading order,
the advective contribution to the average diffusive mass flux is approximated
using the solution of the \emph{linearized} equations, \[
-\av{\left(\delta c\right)\left(\delta\V v\right)}\approx-\av{\left(\delta c\right)\left(\delta\V v\right)}_{\text{linear}}=\left(\D{\chi}\right)\grad\bar{c}.\]
The \emph{effective }diffusion coefficient $\chi_{\text{eff}}=\chi+\D{\chi}$
thus includes an \emph{enhancement} $\D{\chi}$ due to thermal velocity
fluctuations, in addition to the \emph{bare }diffusion coefficient
$\chi$.

The solution to the linearized form of (\ref{eq:LLNS_incomp_v},\ref{eq:LLNS_incomp_c})
in the Fourier domain \citet{ExtraDiffusion_Vailati,DiffusionRenormalization}
shows that the concentration fluctuations and the fluctuations of
velocity parallel to the gradient develop long ranged correlations,
\begin{equation}
\mathcal{S}_{c,v_{\Vert}}=\av{(\widehat{\delta c})(\widehat{\d v}_{\Vert}^{\star})}=-\frac{k_{B}T}{\rho(\nu+\chi)k^{2}}\left(\sin^{2}\theta\right)\,\nabla\bar{c}.\label{eq:S_c_vy}\end{equation}
where $\theta$ is the angle between $\V k$ and $\grad\bar{c}$,
$\sin^{2}\theta=k_{\perp}^{2}/k^{2}$, a hat denotes the Fourier transform,
and star denotes the complex conjugate. The power-law divergence for
small $k$ indicates long ranged correlations between $\delta c$
and $\d{v_{\parallel}}$, and is the cause of both the giant fluctuation
phenomenon and the diffusion enhancement. As seen from (\ref{eq:av_j1}),
the actual correlation that determines the diffusion enhancement is
$\mathcal{S}_{\rho_{1},v_{\Vert}^{(1)}}=\av{(\widehat{\delta c})(\widehat{\d v}_{\Vert}^{(1)})^{\star}}\approx\bar{\rho}\mathcal{S}_{c,v_{\Vert}}$.

We verify the predictions of fluctuating hydrodynamics by using the
Direct Simulation Monte Carlo (DSMC) particle algorithm \citet{DSMC_Bird}.
Previous careful measurements of transport coefficients in DSMC have
been limited to quasi one-dimensional simulations \citet{DSMCConductivity_Gallis}.
The effect we are exploring here does not appear in one dimension
as it arises because of the presence of vortical modes in the fluctuating
velocities. We have performed DSMC calculations for an ideal hard-sphere
gas with molecular diameter $\sigma=1$ and molecular mass $m=1$,
at an equilibrium density of $\rho_{0}=0.06$, with the temperature
kept at $k_{B}T_{0}=T_{0}=1$ via thermal collisions with the top
and bottom walls. Each DSMC particle represents a single hard sphere
so the mean free path is $\lambda=3.75$ and the mean free collision
time is $\tau=2.35$. The DSMC time step was chosen to be $\D t=\tau/2$,
and the collision cell size is either $\D x_{c}=\lambda$ or $\D x_{c}=2\lambda$.
A uniform concentration gradient along the vertical ($y$) direction
is implemented by randomly selecting the species of particles to be
one with probability $c_{T/B}$ if they collide with the top/bottom
wall. Hydrodynamic quantities such as velocity and concentration are
calculated from the particle data by using a grid of $N_{x}\times N_{y}\times N_{z}$
\emph{sampling} or \emph{hydrodynamic cells}, each of volume $\D V=\D x\,\D y\,\D z$,
and a discrete Fourier transform is used to obtain static structure
factors.

\begin{figure}[tbph]
\begin{centering}
\includegraphics[width=0.95\columnwidth]{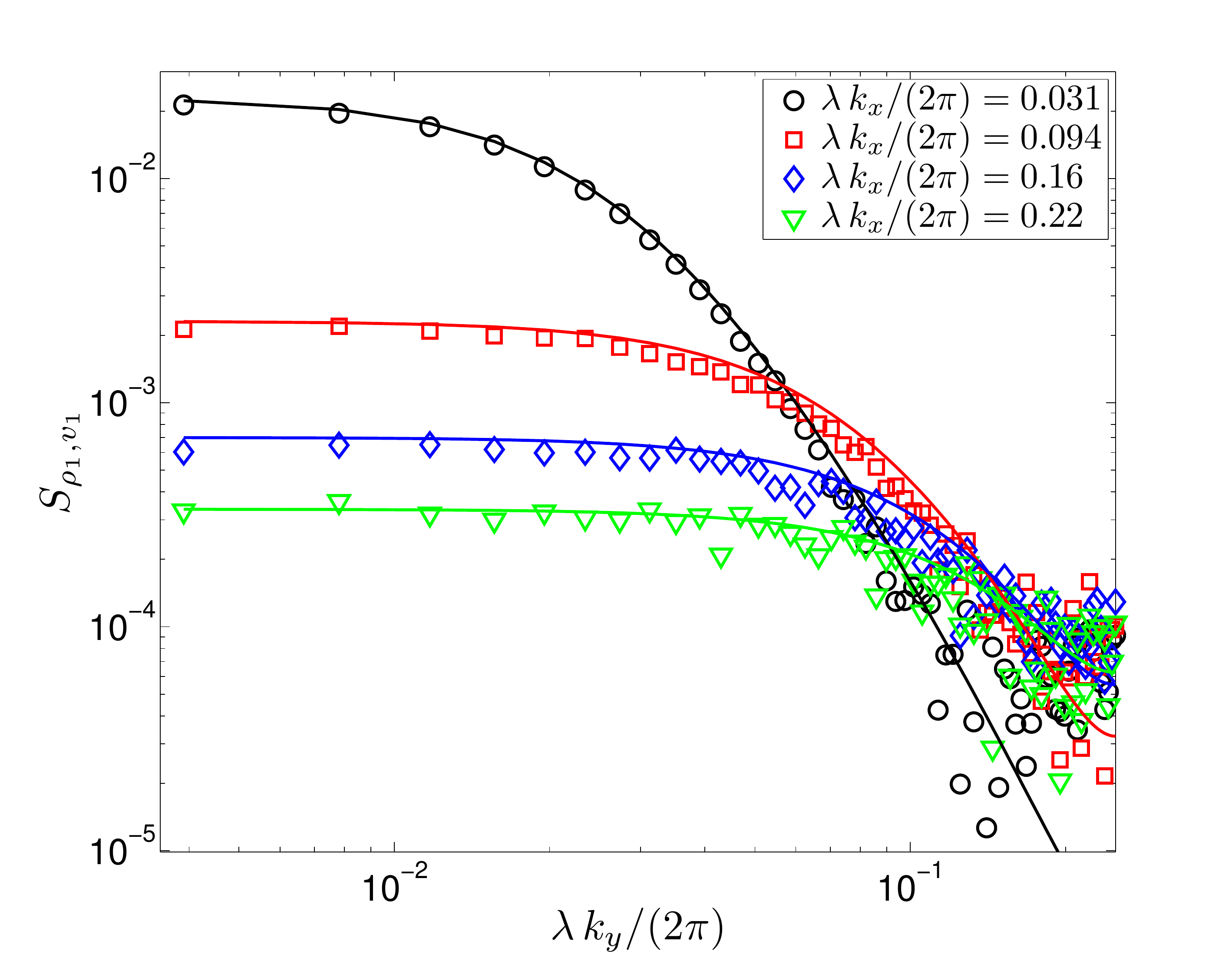}
\par\end{centering}

\caption{\label{fig:DSMC_S_rho1v1}(Color online) Discrete structure factor
$\mathcal{S}_{\rho_{1},v_{\Vert}^{(1)}}$ from quasi two-dimensional
DSMC runs with $L_{x}=64\lambda$, $L_{y}=512\lambda$ and $L_{z}=2\lambda$,
for several wavenumbers $k_{x}=\kappa_{x}\cdot2\pi/L_{x}$ (see legend),
compared to the discrete equivalent of the continuum prediction (\ref{eq:S_c_vy})
(solid lines of the same color). Note that for a fixed $k_{x}$ we
expect the structure factor to decay as $k_{y}^{-4}$.}

\end{figure}

To compare the prediction (\ref{eq:S_c_vy}) to results from particle
simulations, we have converted the \emph{continuum} static structure
factor $\mathcal{S}_{c,v_{\Vert}}(\V k)$ into a \emph{discrete} structure
factor $S_{c,v_{\Vert}}(\V{\kappa})$ for finite-volume averages of
the continuum fields, where the wavenumbers $\V{\kappa}\in\Set Z^{3}$
index the discrete set of wavevectors compatible with periodicity
\citet{DiffusionRenormalization}. In Fig. \ref{fig:DSMC_S_rho1v1}
we compare the theoretical prediction for $\bar{\rho}S_{c,v_{\Vert}}(\V{\kappa})$
to DSMC results for the discrete structure factor $\mathcal{S}_{\rho_{1},v_{\Vert}^{(1)}}$.
Similar results are obtained for two different sizes of the DSMC collision
cells \citet{DiffusionRenormalization}, $\D x_{c}=2\lambda$ and
$\D x_{c}=\lambda$, verifying that the details of the microscopic
collision dynamics do not affect the mesoscopic hydrodynamic behavior.

It is expected that compressibility effects would affect $\mathcal{S}_{\rho_{1},v_{\Vert}^{(1)}}$.
In order to construct a theoretical prediction, however, one must
not only include the effect of compressibility but also replace the
{}``one-fluid'' approximation with a corresponding {}``two-fluid''
compressible hydrodynamic theory \citet{BinaryMixKineticTheory}.
As Fig. \ref{fig:DSMC_S_rho1v1} demonstrates, the incompressible
isothermal theory for $\bar{\rho}S_{c,v_{\Vert}}$ can be used as
a \emph{proxy} for $\mathcal{S}_{\rho_{1},v_{\Vert}^{(1)}}$ in order
to construct theoretical predictions for the diffusion enhancement.

The mass flux due to advection by the fluctuating velocities can be
approximated using (\ref{eq:S_c_vy}) as\begin{equation}
\av{\left(\delta c\right)\left(\delta\V v\right)}_{\text{linear}}=\left(2\pi\right)^{-3}\int_{\V k}\mathcal{S}_{c,\V v}\left(\V k\right)\, d\V k,\label{eq:dcdv_realspace}\end{equation}
giving an estimate of the diffusion enhancement \citet{ExtraDiffusion_Vailati,DiffusionRenormalization}\begin{equation}
\D{\chi}=\frac{k_{B}T}{(2\pi)^{3}\rho\left(\chi+\nu\right)}\;\int_{\V k}\left(\sin^{2}\theta\right)k^{-2}\, d\V k.\label{eq:dchi_k_int}\end{equation}
Because of the $k^{-2}$-like behavior, the integral over all $\V k$
above diverges unless one imposes \citet{ExtraDiffusion_Vailati}
a lower bound, $k_{\min}\sim2\pi/L$ in the absence of gravity, \emph{and}
a phenomenological cutoff $k_{\max}\sim\pi/L_{\text{mol}}$ for the
upper bound, where $L_{\text{mol}}$ is an ad-hoc {}``molecular''
length scale.

For a quasi two-dimensional system, $L_{z}\ll L_{x}\ll L_{y}$, we
can replace the integral over $k_{z}$ with $2\pi/L_{z}$ and integrate
over all $k_{y}$. This leads to an average total diffusive flux that
grows logarithmically with the width $L_{x}$ for a fixed height $L_{y}$
\citet{DiffusionRenormalization}, \begin{equation}
\chi_{\text{eff}}^{(2D)}\approx\chi+\frac{k_{B}T}{4\pi\rho(\chi+\nu)L_{z}}\,\ln\frac{L_{x}}{L_{0}},\label{eq:chi_eff_2D}\end{equation}
where $L_{0}>2L_{\text{mol}}$ is the \emph{reference} width at which
the {}``bare'' diffusion coefficient is measured. Within the phenomenological
perturbative theory $L_{0}$ is an arbitrary (mesoscopic) length scale,
and simply defines $\chi=\chi_{\text{eff}}(L_{x}=L_{0})$. For comparison
between the particle simulations and the theory we use $L_{0}=16\lambda$.
When the system width becomes comparable to the height, boundaries
will intervene and for $L_{x}\gg L_{y}$ the effective diffusion coefficient
must become a constant, which is predicted to be a logarithmically-growing
function of $L_{y}$ in two dimensions. The same coefficient in front
of $\ln L_{x}$ as in (\ref{eq:chi_eff_2D}) is obtained when the
integral over $k_{x}$ is replaced by a discrete sum over the wavenumbers
consistent with periodicity, $k_{x}=\kappa_{x}\cdot2\pi/L_{x}$, $\kappa_{x}\in\Set Z$
\citet{DiffusionRenormalization}.

In three dimensions, $L_{x}=L_{z}=L\ll L_{y}$, $\chi_{\text{eff}}$
converges as $L\rightarrow\infty$ to the macroscopic diffusion coefficient,
\begin{equation}
\chi_{\text{eff}}^{(3D)}\approx\chi+\frac{\alpha\, k_{B}T}{\rho(\chi+\nu)}\left(\frac{1}{L_{0}}-\frac{1}{L}\right),\label{eq:chi_eff_3D}\end{equation}
but for a finite system the effective diffusion coefficient is reduced
by an amount $\sim L^{-1}$ due to the truncation of the velocity
fluctuations by the confining walls. Calculating the exact value of
$\alpha$ requires performing a sum over $\kappa_{x}$ and $\kappa_{z}$
instead of integrals over $k_{x}$ and $k_{z}$, as we have done numerically
\citet{DiffusionRenormalization}. The numerical results suggest that,
as in two dimensions, the difference in $\chi_{\text{eff}}^{(3D)}$
between two systems attains a finite value as $L_{\text{mol}}\rightarrow0$,
justifying (\ref{eq:chi_eff_3D}) for $\left(L_{0},L\right)\gg L_{\text{mol}}$.

In particle simulations, we calculate the \emph{effective} diffusion
coefficient $\chi_{\text{eff}}$ from the momentum density of one
of the species (denoted either with a subscript or with a parenthesis
superscript) along the vertical direction,\begin{equation}
\av{j_{\parallel}^{(1)}}=\av{\rho_{1}v_{\Vert}^{(1)}}=\rho_{0}\chi_{\text{eff}}\,\frac{\bar{c}_{T}-\bar{c}_{B}}{L_{y}-\D y}\approx\rho_{0}\chi_{\text{eff}}\left(\nabla\bar{c}\right),\label{eq:eff_definition}\end{equation}
where we measure $\bar{c}_{T}$ and $\bar{c}_{B}$ in the top and
bottom layer of sampling cells (whose centers are a distance $L_{y}-\D y$
from each other) to empirically account for the small concentration
slip in DSMC. Numerical experiments have verified that $\av{j_{\parallel}^{(1)}}$
matches the flux obtained from counting the average number of color
flips at the top or bottom walls. Furthermore, the results verify
that $\av{j_{\parallel}^{(1)}}$ is linear in the gradient $\nabla\bar{c}$,
and that $\bar{c}_{T/B}$ are essentially independent of the system
dimensions.

The traditional definition of a {}``renormalized'' diffusion coefficient
\citet{DiffusionRenormalization_I} as the macroscopic limit of $\chi_{\text{eff}}$,
only works in three dimensions and is not very useful for confined
systems. Instead, for each sampling cell, we define a \emph{locally
renormalized} diffusion coefficient $\chi_{0}$ via\begin{equation}
\av{\rho_{1}}\av{v_{\Vert}^{(1)}}=\av{\rho_{1}}\av{j_{\parallel}^{(1)}/\rho_{1}}=\bar{\rho}\chi_{0}\left(\nabla\bar{c}\right),\label{eq:bare_definition}\end{equation}
where we have accounted for the fact that the macroscopic concentration
profile $\bar{c}(y)$ may depend on $y$. In fact, such a dependence
is observed in the particle simulations, and we have approximated
the local concentration gradient $d\bar{c}/dy$ by a numerical derivative
of a polynomial of degree five fit to $\bar{c}(y)$. We have empirically
observed that $\chi_{0}$ is independent of $y$, except for a boundary
layer close to the top and bottom walls \citet{DiffusionRenormalization}.
This is an important finding, since (\ref{eq:bare_definition}) is
a constitutive model that is assumed to hold not just at the macroscale
but also at the mesoscale, notably, $\chi_{0}$ is an input parameter
for fluctuating hydrodynamics finite-volume solvers \citet{LLNS_S_k}.

\begin{figure}[tbph]
\begin{centering}
\includegraphics[width=0.99\columnwidth]{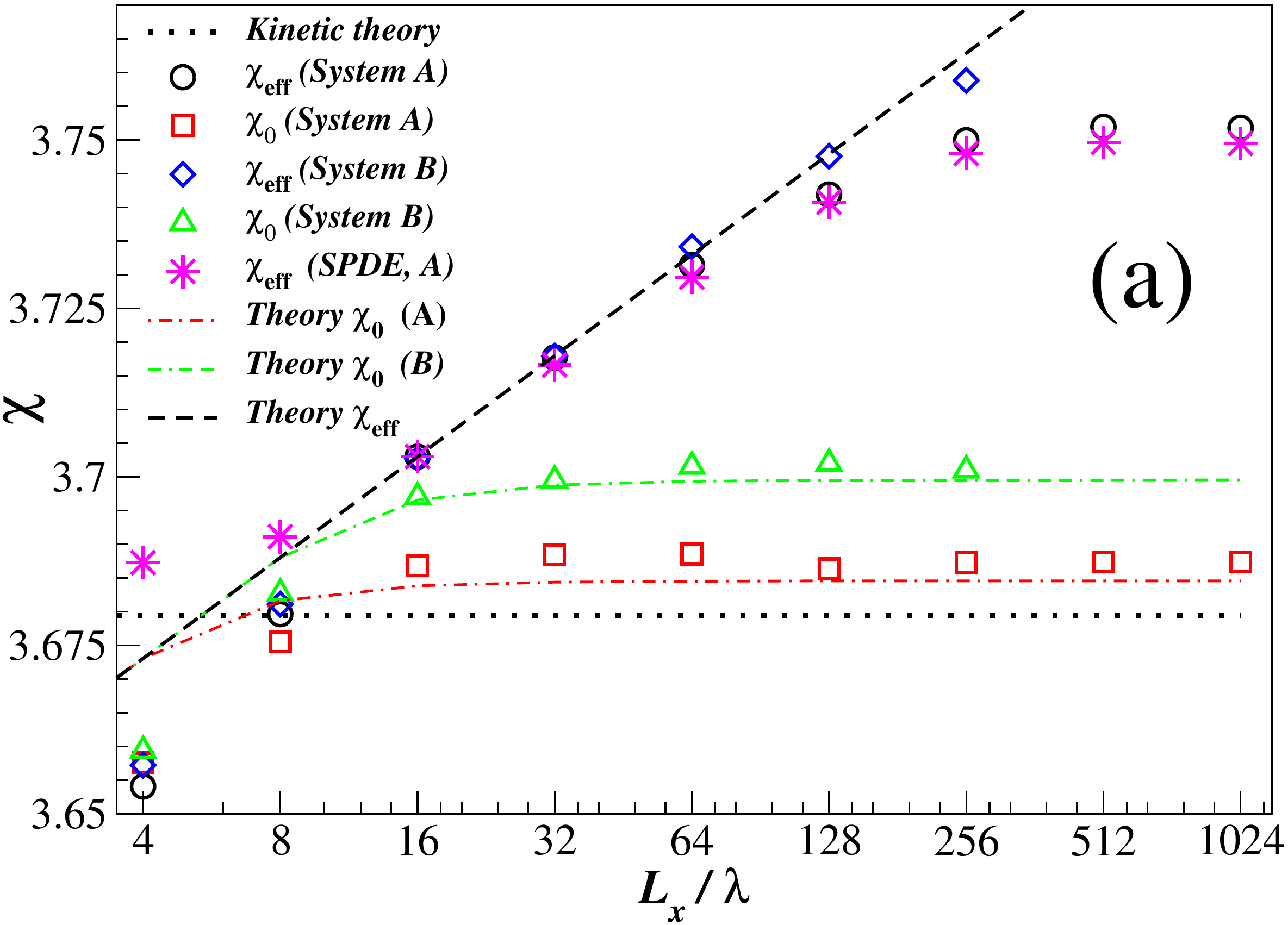}
\par\end{centering}

\begin{centering}
\includegraphics[width=0.99\columnwidth]{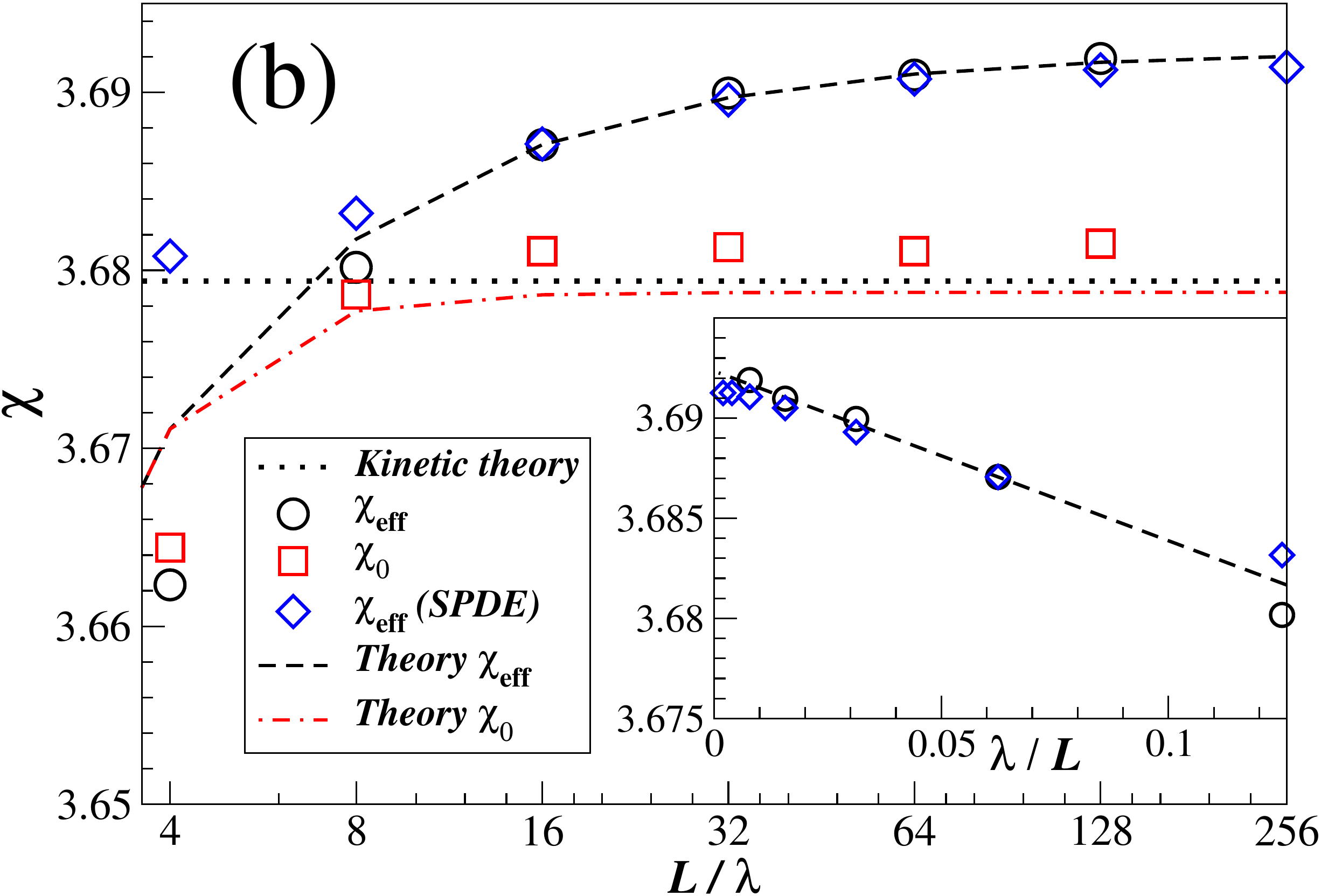}
\par\end{centering}

\caption{\label{fig:j1_vs_Nx}(Color online) (\emph{Panel a}) The effective
$\chi_{\text{eff}}$ and the renormalized $\chi_{0}$ diffusion coefficients
as a function of the width of the system $L_{x}$ in two dimensions.
Numerical results for System A (DSMC and SPDE) and System B (DSMC)
are shown with symbols (see legend). The error bars for all of the
simulation data are comparable or smaller than the size of the symbols.
The theoretical predictions \citet{DiffusionRenormalization} are
evaluated numerically and shown with lines. (\emph{Panel b}) Same
as panel (a) but in three dimensions. The inset highlights the $L^{-1}$
behavior.}

\end{figure}

Figure \ref{fig:j1_vs_Nx}a shows how $\chi_{\text{eff}}$ and $\chi_{0}$
change as the width of the system $L_{x}$ is increased while keeping
the height $L_{y}$ fixed for two different quasi two-dimensional
DSMC systems. For System A, the DSMC collision cells are cubes of
side $\D x_{c}=7.5=2\lambda$, while each sampling cell contains $2\times2\times1$
collision cells, or $N_{p}=101$ particles on average. The height
of the box is $L_{y}=256\lambda=960$ and the imposed concentrations
at the walls are $c_{B}=0.25$ and $c_{T}=0.75$. For System B, the
sampling cells are twice as large, $4\times4\times1$ collision cells
each, and the system height is $L_{y}=512\lambda=1920$. We obtain
similar results using a factor of two smaller collision cells (not
shown). For the quasi two-dimensional systems, the thickness is $L_{z}=7.5=2\lambda$
and there is only one DSMC collision cell along the $z$ direction.
Figure \ref{fig:j1_vs_Nx}a shows that $\chi_{\text{eff}}$ grows
like $\ln L_{x}$, with a slope that is well-predicted by Eq. (\ref{eq:chi_eff_2D}).
For widths larger than about $8$ mean free paths, $\chi_{0}$ becomes
constant and rather similar to the Chapman-Enskog kinetic theory prediction.
Note that $\chi_{0}$ is not a fundamental material constant and in
fact depends on the shape of the sampling cells, notably, it grows
as the sampling cell size is enlarged.

In Fig. \ref{fig:j1_vs_Nx}b we show results from three dimensional
DSMC simulations, in which the system width ($x$) and depth ($z$)
directions are equivalent, $L_{z}=L_{x}=L$, and the rest of the parameters
are as for System A. Similar behavior is seen as in two dimensions,
except that now the effective diffusion grows as $-L^{-1}$ and saturates
to a constant value for large $L$, assuming that $L_{y}\gg L$.

The predictions of the simplified fluctuating hydrodynamic theory,
Eqs. (\ref{eq:chi_eff_2D}) and (\ref{eq:chi_eff_3D}), are shown
in Figs. \ref{fig:j1_vs_Nx} and are seen to be in very good agreement
with the particle simulations for intermediate $L_{x}$. However,
recall that the incompressible isothermal theory assumed that $L_{y}$
is essentially infinite and thus in two dimensions $\chi_{\text{eff}}$
grows unbounded in the macroscopic limit. Yet when $L_{x}\gg L_{y}$,
$\chi_{\text{eff}}$ must saturate to a constant value, and the particle
data shown in Fig. \ref{fig:j1_vs_Nx}a shows measurable deviations
from the simple theory for $L_{x}\gtrsim L_{y}/2$. One can extend
the theoretical calculations to account for the hard wall boundary
conditions in the $y$ direction \citet{FluctHydroNonEq_Book}, however,
such a calculation is non trivial. Instead, we have used the finite-volume
solver developed in Ref. \citet{LLNS_S_k} to solve the LLNS non-linear
system of SPDEs for the same system dimensions as in the particle
simulations. To minimize the effect of nonlinearities in the SPDE
solver, we artificially reduce the amplitude of the noise by some
factor $\epsilon\ll1$, but then scale all correlations by $\epsilon^{-2}$
\citet{DiffusionRenormalization}. The results, shown in Fig. \ref{fig:j1_vs_Nx},
are in excellent agreement with the particle simulations for the larger
system sizes.

In finite-volume solvers, the spacing of the computational grid plays
the equivalent of the cutoff length $L_{\text{mol}}$, and therefore
$\chi_{\text{eff}}$ depends on the grid spacing. We have added a
constant to the effective diffusion coefficient obtained from SPDE
runs so as to match $\chi_{\text{eff}}$ from the particle simulations
for $L_{x}=L_{0}=16\lambda$. This correction essentially renormalizes
$\chi_{0}$ based on the size of the finite-volume hydrodynamic cells.
One can think of $\chi_{0}$ as the physical-space equivalent of the
wavenumber-dependent diffusion coefficient $\chi\left(\V k,\omega=0\right)$
commonly used in linear response theories \citet{DiffusionRenormalization_I,SelfDiffusion_Linearity}.
Theoretical predictions \citet{DiffusionRenormalization} for $\chi_{0}$
indicate that $\chi_{0}$ only includes {}``sub-grid'' contributions,
from wavenumbers larger than $2\pi/\D x$. Thus $\chi_{0}$ stops
increasing once the system becomes substantially larger than the size
of the sampling cell. The bare diffusion coefficient $\chi$ in the
theory and SPDE calculations is adjusted so that for $L_{x}=L_{0}=16\lambda$
the effective diffusion is the same as that measured in the particle
simulations.

Previously-studied corrections to the bare or molecular transport
coefficients due to the tail of the velocity autocorrelation function
(VACF) \citet{LongRangeCorrelations_MD}, hydrodynamic interactions
with periodic images of a given particle \citet{FiniteSize_Diffusion_MD},
and the contribution due to advection by thermal velocity fluctuations
\citet{DiffusionRenormalization_I,ExtraDiffusion_Vailati} studied
here, are all the same physical phenomenon simply calculated through
different theoretical approaches, all of which are \emph{equivalent}
because of linearity \citet{DiffusionRenormalization} . In three
dimensions, simple estimates indicate that the contribution of fluctuations
to the macroscopic diffusion coefficient are small compared to molecular
effects for gases but can be significant for liquids \citet{DiffusionRenormalization}.
However, the logarithmic divergence in (\ref{eq:chi_eff_2D}) means
that $\D{\chi}\gg\chi$ for sufficiently large (quasi) two-dimensional
systems, requiring the inclusion of higher order corrections in the
theory. At present, reaching the system width $L_{x}$ where $\D{\chi}\sim\chi$
is difficult with DSMC simulations, but it may be accessible to finite-volume
SPDE solvers or experiments \citet{GiantFluctuations_ThinFilms}.

Our results conclusively demonstrate that the advection by thermal
velocity fluctuations affects the \emph{mean} transport in nonequilibrium
finite systems. Theoretical modeling of finite systems at the nano
or microscale thus requires including nonlinear hydrodynamic fluctuations.
The advantage of fluctuating hydrodynamics is that it is simple, and
it can take into account the proper boundary conditions and exact
geometry, especially if a numerical SPDE solver is used. Furthermore,
other effects such as gravity, temperature variations, or time dependence,
can easily be included. However, a proper fully-nonlinear theory has
yet to be developed, and requires detailed understanding of the role
of the necessary large wavenumber cutoffs (regularizations). Future
work should verify the predictions of fluctuating hydrodynamics for
the effect of fluctuations on diffusive transport in spatially non-uniform
systems.

We thank Berni Alder, Doriano Brogioli, Jonathan Goodman and Eric
Vanden-Eijnden for informative discussions. This work was supported
by the DOE Applied Mathematics Program (DE-AC02-05CH11231).

\end{document}